\documentclass{iaus}
\usepackage{graphicx}

\newcommand{\FeI}{Fe\,{\sc i}}

\newcommand\aap{A\&A}
\newcommand\apj{ApJ}  
\newcommand\solphys{Sol.~Phys.}
\newcommand\apjl{ApJ}
\newcommand\nat{Nature}
\usepackage{natbib} 
\title[JD 11.~~What determines the penumbral size and Evershed flow speed?] {What determines the penumbral size and Evershed flow speed?}

\author[Deng et al.]   {Na Deng$^{1,2}$, Toshifumi Shimizu$^{3}$, Debi Prasad Choudhary$^{1}$ \and Haimin Wang$^2$}

\affiliation{$^1$Physics and Astronomy Department, California State University Northridge, \\ 18111 Nordhoff St., Northridge, CA 91330, United States \\ email: {\tt na.deng@csun.edu} \\[\affilskip]
$^2$Space Weather Research Laboratory, New Jersey Institute of Technology, \\ 323 Martin Luther King Blvd., Newark, NJ 07102, United States \\[\affilskip]
$^3$Institute of Space and Astronautical Science, Japan Aerospace Exploration Agency, \\ 3-1-1 Yoshinodai, Sagamihara, Kanagawa 229-8510, Japan}

\pubyear{2011}
\volume{273}  \pagerange{1--5}
\jname{Physics of Sun and Star Spots}
\editors{A.C. Editor, B.D. Editor \& C.E. Editor, eds.}
\begin{document}

\maketitle

\begin{abstract}
Using Hinode SP and G-band observations, we examined the relationship between magnetic field structure and penumbral size as well as Evershed flow speed. The latter two are positively correlated with magnetic inclination angle or horizontal field strength within 1.5 kilogauss, which is in agreement with recent magnetoconvective simulations of Evershed effect. This work thus provides direct observational evidence supporting the magnetoconvection nature of penumbral structure and Evershed flow in the presence of strong and inclined magnetic field.
\end{abstract}

\firstsection               \section{Introduction and Motivation}
The penumbra along with the coupled Evershed flow has been one of the most intriguing phenomena on the Sun that motivated many detailed observational and theoretical studies. High resolution imaging and spectro-polarimetric analyses have revealed that the penumbra consists of two distinct magnetic components. The more inclined magnetic component (60$^\circ$--100$^\circ$ from inner to outer penumbra, with respect to the surface normal) with weaker field strength ($\thicksim$1200 Gauss) is embedded in the less inclined magnetic background (30$^\circ$-- 60$^\circ$ from inner to outer penumbra) that has stronger field strength ($\thicksim$1700 Gauss), which is frequently referred as ``uncombed'' structure \citep[][and references therein]{Solanki+Montavon1993A&A...275..283S, Langhans+etal2005A&A...436.1087L, Beck2008A&A...480..825B, Borrero2009ScChG..52.1670B}. The Evershed flow is magnetized and mainly carried by the more inclined magnetic component \citep[e.g.,][]{title+etal1993, stanchfield+thomas+lites1997, BellotRubio+etal2004A&A...427..319B, Borrero+etal2005A&A...436..333B, Ichimoto+etal2008A&A...481L...9I, Deng+etal2010ApJ...719..385D}.

Several theoretical models have been proposed to understand the penumbral structure and the mechanism of the Evershed flow, such as siphon-flow with downward pumping of magnetic flux \citep{Montesinos+Thomas1997Natur.390..485M, Thomas+etal2002Natur.420..390T}, embedded moving flux tube model \citep{schlichenmaier+jahn+schmidt1998a}, ``gappy
penumbra'' model \citep{Scharmer+Spruit2006A&A...460..605S}, and elongated convective granular cells by the presence of inclined strong magnetic field \citep{Hurlburt+etal1996ApJ...457..933H, Hurlburt+etal2000SoPh..192..109H}. In particular, recent realistic three-dimensional numerical MHD simulations have successfully reproduced the penumbral structure and Evershed flow as a natural consequence of thermal magnetoconvection when the average inclination of kilogauss magnetic field is larger than 45$^\circ$ \citep{Rempel+etal2009Sci...325..171R}. Moreover, the filamentary patten and the speed of the simulated Evershed horizontal surface flow are strongly controlled by both strength and inclination of the magnetic field with inclination playing most important role \citep{Rempel+etal2009ApJ...691..640R, Kitiashvili+etal2009ApJ...700L.178K}.

On the other hand, rapid penumbral decay (disappearance or length reduction) and change of associated Evershed flow were found right after major flares in the periphery of complex $\delta$ sunspots \citep{WangH+etal2004ApJ...601L.195W, LiuC+etal2005ApJ...622..722L, Deng+etal2005ApJ...623.1195D}. The authors interpret the sudden change of penumbral white light structure to be a result of the change of overall magnetic field inclination down to the photosphere due to magnetic reconnection in the flare, which was then confirmed by other authors \citep{Sudol+Harvey2005ApJ...635..647S, Li+etal2009ScChG..52.1702L}.

Both aforementioned magnetoconvective simulations and observations during flares hint a relationship between magnetic field structure (especially inclination) and penumbral size as well as Evershed flow speed. While a systematic examination of such relationship from real observation is still missing. Thanks to the high quality measurement of vector magnetic field by Hinode, we investigate how the magnetic parameters are related to the penumbral size and Evershed flow speed by analyzing sunspots of different sizes (i.e., in large scale) and the properties in different sectors within a sunspot (i.e., in small scale).

\begin{figure}[t]
\begin{center}
\includegraphics[width=5.2in]{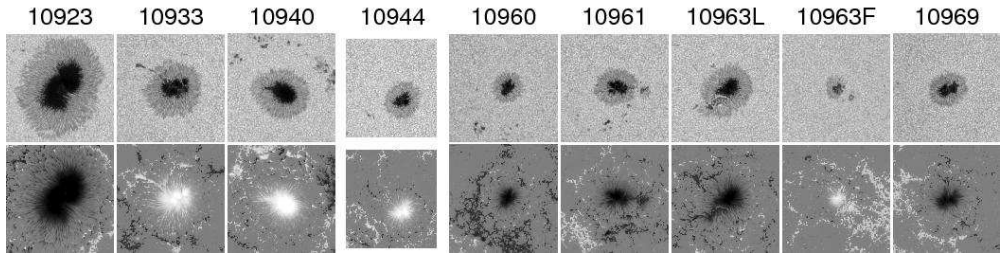}
\caption{Hinode/SP \FeI\ 630 nm continuum images (top row) and B$_z$ component magnetograms (bottom row) of the 9 sunspots with their NOAA AR numbers labeled on top. The FOV for each panel is $90^{\prime\prime} \times 90^{\prime\prime}$.}
\label{fig1}
\end{center}
\end{figure}

\begin{figure}
\begin{center}
\includegraphics[width=5.2in]{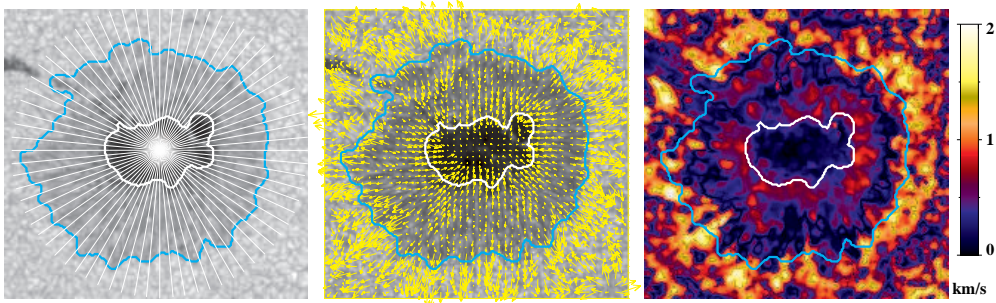}
\caption{The SP continuum image, G-band LCT flow map and flow speed map of NOAA 10933. The sunspot is evenly divided into 360 sectors from the Center-Of-Mass of the umbra.}
\label{fig2}
\end{center}
\end{figure}

\section{Observation and Data Reduction}
We studied 9 simple $\alpha$ sunspots at the late phase of solar cycle 23. They are close to disk center (heliocentric angle $<12^\circ$) and exhibit different size. Fig.\,\ref{fig1} shows their continuum images and B$_z$ magnetograms obtained by Hinode Spectropolarimeter (SP). From the 13 parameters generated by a Milne-Eddington Stoke inversion of the SP data, we used or calculated the following most relevant parameters: continuum intensity, magnetic inclination angle, horizontal field strength, and field strength. The 180$^\circ$ azimuthal ambiguity of the inverted magnetic field was resolved using the ``minimum energy'' algorithm \citep{Metcalf1994SoPh..155..235M}. We transformed the measured magnetic field vectors to local Cartesian coordinates so that the inclination is with respect to the surface normal. To measure Evershed flow, we used Local Correlation Tracking (LCT) technique based on a 1hr series of Hinode G-band images (2min cadence) co-aligned and co-temporal with SP data. Same tracking window size was used for all the sunspots. Fig.\,\ref{fig2} shows the continuum intensity, LCT flow map and flow speed map of a sunspot.

The continuum images were smoothed and contoured to outline the penumbral areas, whose inner and outer boundaries are about 0.45$I_0$ and 0.9$I_0$ ($I_0$ is the quiet Sun continuum intensity), respectively. The penumbral lengthes were measured in each sector (i.e., the distance between the inner and outer boundaries, see panel a of Fig.\,\ref{fig2}). The aforementioned magnetic parameters and LCT outward flow speed were also averaged over each sector's area. All these quantities were then averaged over the entire penumbral area for each sunspot. Sectors where the penumbral structure is complex or deviate from radial direction much were excluded from consideration.

\begin{figure}
\begin{center}
 \includegraphics[width=5.2in]{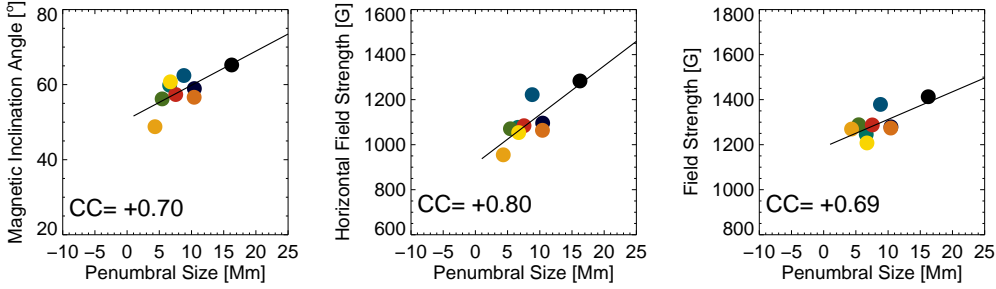}
 \caption{Scatter plots between penumbral length and magnetic parameters averaged over the whole penumbral areas of the 9 sunspots. The linear correlation coefficient (CC) are labeled.}
   \label{fig3}
\end{center}
\end{figure}

\begin{figure}
\begin{center}
 \includegraphics[width=5.2in]{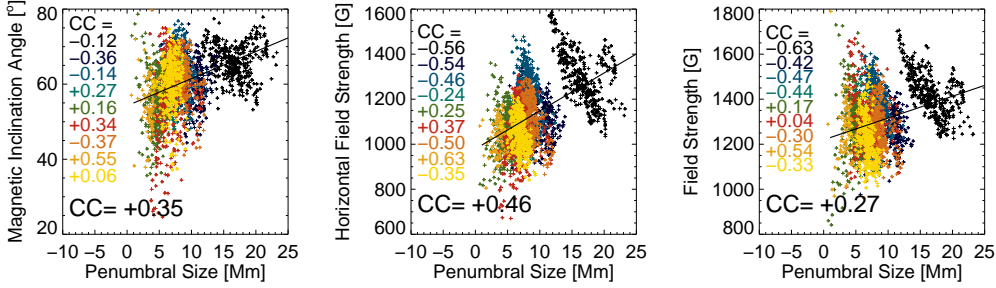}
 \caption{Scatter plots between penumbral length and magnetic parameters averaged in each sectors for the 9 sunspots. Each spot is represented by one color. The CC for each spot and for all the data points are labeled.}
   \label{fig4}
\end{center}
\end{figure}

\begin{figure}
\begin{center}
 \includegraphics[width=5.2in]{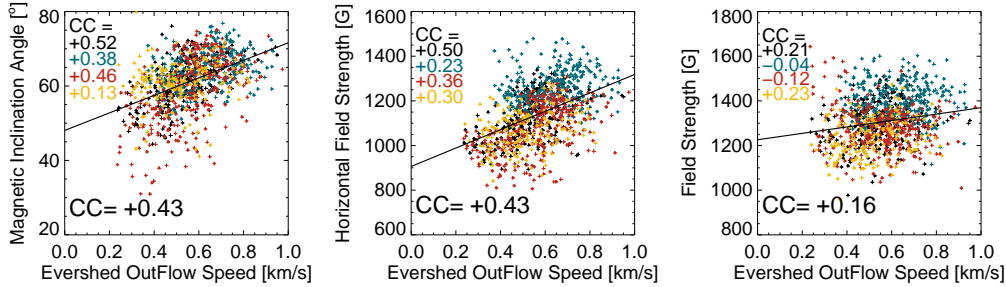}
 \caption{Scatter plots between outward Evershed flow speed and magnetic parameters averaged in each sectors for 4 sunspots that have co-temporal SP and G-band data.}
   \label{fig5}
\end{center}
\end{figure}

\section{Results and Conclusion}
Fig.\,\ref{fig3} shows that in large scale the penumbral size is well correlated with magnetic inclination angle and horizontal field strength. The mean inclination angles in penumbrae are all greater than 45$^\circ$ for the 9 sunspots. Fig.\,\ref{fig4} plots the same but in small scale. While all the data points still follow the same trend as in Fig.\,\ref{fig3}, for each individual sunspot the trend is not consistent. This might be due to dynamic and chaotic fluctuations in small scale or over simplification of our method. Fig.\,\ref{fig5} shows that the Evershed flow speed is always positively correlated with magnetic inclination angle and horizontal field strength in both small and large scales for the 4 sunspots studied. This result is consistent with magnetoconvective simulations of Evershed effect, where the horizontal flow speed increases with larger inclination angle within certain field strength range \citep{Kitiashvili+etal2009ApJ...700L.178K}. This work thus provides direct observational evidence supporting the magnetoconvection nature of penumbral structure and Evershed flow under strong and inclined magnetic field.

\acknowledgments

N.D. and D.P.C. were supported by NASA grant NNX08AQ32G and NSF grant ATM 05-48260. N.D. thank valuable discussions with Drs. Ichimoto, Kitiashvili, Martinez Pillet, Sainz Dalda, and Schlichenmaier.

\end{document}